\documentclass[prl,twocolumn,nofootinbib]{revtex4}
\usepackage{dcolumn}
\usepackage{amssymb}
\usepackage{bm}
\usepackage{enumerate}
\def\be{\begin{enumerate}}                     \def\ee{\end{enumerate}}
\def\beq{\begin{equation}}                     \def\eeq{\end{equation}}
\def\bea{\begin{eqnarray}}   \def\eea{\end{eqnarray}}

\def\3halfs{\textstyle{\frac{3}{2}}}                       

\def\ben{\begin{enumerate}}                   \def\een{\end{enumerate}}
\def\bitem{\begin{itemize}} \def\eitem{\end{itemize}}

         \def\etal{{\emph{et         al}}}

\begin{document}
\def\Universita{Universit\`a}
\title{A Note on High Energy Neutrinos from AGN Cores}
\author{F.W. Stecker} \affiliation{NASA Goddard Space Flight Center}
\affiliation{Greenbelt, MD 20771, USA}

\begin{abstract}

Taking  account of  new physics  and astronomy  developments I  give a
revised high energy  neutrino flux for the AGN  core model of Stecker,
Done, Salamon and Sommers.
\end{abstract}
\maketitle
In 1991 we proposed a model suggesting that very high energy neutrinos
could  be produced in  the cores  of active  galaxies (AGN) such as Seyfert
galaxies  \cite{sdss91}. Using that  model, we  gave estimates  of the
flux  and  spectrum  of  high  energy neutrinos  to  be  expected (see
\cite{note}.)

The  fluxes given in  Ref. \cite{sdss91}  were normalized  by assuming
that 100\%  of the  X-ray background was  nonthermal radiation  from a
superposition of unresolved  Seyfert galaxies. Subsequent observations
of these AGN  have shown that their emission  is predominantly thermal
and therefore can not be  directly related to the production of highly
relativistic particles in such sources.

However,  the extragalactic background  radiation at  MeV $\gamma$-ray
energies may  be due to nonthermal  $\gamma$-ray emission from
AGN.  {\it COMPTEL} observations of the galactic black hole source Cyg
X-1,  show a  hard tail  of emission,  extending out  to  MeV energies
\cite{mcc97}.  An  explanation that has been suggested  to account for
this MeV tail is  that the electron  distribution is  not completely
thermalized  \cite{pc98}.  This  is  physically reasonable  since  the
thermalization timescales  for the electrons  can be smaller  than the
other  timescales  in  these  systems (e.g.  Ref.   \cite{cop99})  The
overall 2 keV to 5 MeV spectrum  of Cyg X-1 can then  be modelled if
$90$\%  of the  power  goes into  a  $\sim 100$  keV thermal  electron
distribution,  while the remaining  $\sim 10$\%  is in  the form  of a
nonthermal $\gamma$-ray tail.

It is well known that the low-hard state spectra of Cyg X-1 and other
galactic black hole sources bear  a remarkable similarity to those from
radio quiet AGN  \cite{pou98}, with emission in both  types of sources
involving the same  physical processes of disk accretion  onto a black
hole. Thus,  we expect a  similar hard tail  to be present  in Seyfert
galaxies.  Observations of Seyfert galaxies having flat spectrum radio
nuclei  using the  VLBA have  shown  that these  sources are  emitting
nonthermal radiation from central core regions with sizes $\sim$ 0.05
to 0.2  pc  \cite{mun99}.   Such  cores  may also  be  the  source  of
both nonthermal MeV emission and high energy neutrinos. A tail of the 
Cyg X-1 type  could not  be detected in  an individual AGN  using current
instrumentation, but a  superposition of such tails in  the spectra of
AGN could account  for the reported MeV background  spectrum and flux.
Based on  observations of Cyg X-1, and taking account  of the fact
that AGN contain  much more massive  black holes at their  cores, one
can assume that such sources exhibit a high energy tail of roughly
the same magnitude relative  to the thermal emission as Cygnus X-1.
A superposition of thermal emission from unresolved Seyferts with 
spectra in the X-ray range similar to Cyg X-1 
can reasonably  account  for  the  X-ray  background
\cite{grs99}. A superposition of such AGN spectra with Cyg X-1 type 
tails can  also account for the shape and  flux level of
the MeV  background \cite{ssd99}.
  
In accord with the above arguments, if we assume that 
the extragalactic MeV background is made up of the 10\% component
of nonthermal radiation from Seyferts,
this lowers the estimated AGN core $\nu_{\mu}$ flux  
to an order of magnitude below that previously obtained by normalizing to the
X-ray background in Ref. \cite{sdss91}. A further reduction in 
the $\nu_{\mu}$ flux by a factor of 2 comes from neutrino oscillations, whose
discovery was made after the publication of Ref. \cite{sdss91}.
The new estimate is therefore obtained by lowering the flux shown in
the Figure in the erratum of Ref. \cite{sdss91} by a factor of 20.
This rescaling gives a value for the $\nu_{\mu}$ flux at 100 TeV
of $E_{\nu}^2\Phi (E_{\nu}) \sim 10^{-8}$ GeV cm$^{-2}$s$^{-1}$sr$^{-1}$. 
Such a flux is consistent with present limits from {\it AMANDA} \cite{ach05}.

\noindent This work was supported by NASA Grant ATP03-0000-0057.


\end{document}